\documentclass[10pt, conference, compsocconf]{IEEEtran}
\ifCLASSINFOpdf
\else
\fi

\usepackage[pdftex]{graphicx}
\usepackage[cmex10]{amsmath}

\begin{document}
%
\title{Data Challenges in High-Performance Risk Analytics}


\author{\IEEEauthorblockN{Blesson Varghese\textsuperscript{1} and Andrew Rau-Chaplin}
\IEEEauthorblockA{Risk Analytics Lab, Faculty of Computer Science\\
Dalhousie University, Halifax, Canada\\
Email: \{varghese, arc\}@cs.dal.ca}
}


%


\maketitle
\footnotetext[1]{Corresponding Author}

\begin{abstract}
Risk Analytics is important to quantify, manage and analyse risks from the manufacturing to the financial setting. In this paper, the data challenges in the three stages of the high-performance risk analytics pipeline, namely risk modelling, portfolio risk management and dynamic financial analysis is presented.
\end{abstract}


%
\IEEEpeerreviewmaketitle

\section{Introduction}

Risk analytics \cite{riskanalytics:book} for the reinsurance industry is primarily simulation-based in which large amounts of data are rapidly processed and millions of simulations are quickly performed. This can be achieved only if data is efficiently managed and parallelism is exploited within simulations. Therefore, risk analytics inherently opens an avenue to exploit the synergy that can be obtained by bringing together state-of-the-art approaches in data management and high-performance computing.

In this paper, we briefly discuss the data challenges that need to be addressed in the three stage risk analytics pipeline while employing high-performance computing. The challenges are specific to the nature of data ingested and produced in the different stages of the pipeline. The key observation is that traditional relational databases are of limited use for efficiently implementing the risk analytics pipeline. Approaches in which high-performance computing can be employed for accumulating large distributed memory and large file space for managing data seems the way forward.


\section{Data Challenges in the Risk Analytics Pipeline}
\label{datachallenges}

The risk analytics pipeline is divided into three stages, namely risk modelling, portfolio risk management and dynamic financial analysis. 

In the first stage of the risk analytics pipeline, Catastrophe Models \cite{cat1:book} are employed for quantifying risk. Catastrophe models take two primary inputs, firstly, stochastic event catalogues (i.e., mathematical representations of natural occurrence patterns and characteristics of catastrophes such as earthquakes), and secondly, exposure databases (i.e., description of attributes such as construction type or value of buildings exposed to the catastrophe in a location). An event-exposure pair is analysed using three modules that quantify (i) the hazard intensity at exposure sites, (ii) the vulnerability of the buildings and the resulting damage level, and (iii) the resultant financial loss. The output at this stage is an Event-Loss Table (ELT). The main challenge here lies at the intersection of managing data and high-performance computing. Risk modelling is highly compute and data intensive. Typically, data needs to be organised in a small number of very large tables and streamed by independent processes, further to which the results need to be aggregated.  

An ELT is the risk associated with an individual reinsurance contract, and is the output of the first stage. A reinsurer typically may have tens of thousands of contracts and are interested in quantifying the risk across their whole portfolio requiring a further level of stochastic simulation presenting approximately a million alternate views of a predetermined period, i.e., a contractual year. The data generated in this stage is very large. For example, if an analysis of 10,000 contracts for 100,000 events in 1,000 locations with 50,000 trial years is considered, the Year-Event-Location-Loss Table (YELLT) has over $5 X 10^{16}$ entries. In existing portfolio management tools it is almost impossible to analyse at the YELLT level which highlights the limitation of existing data management strategies employed in such tools.

An additional Monte Carlo simulation, referred to as aggregate analysis \cite{cat4:journal} is necessary for generating an alternate view of which events occur and in which order they occur within a contractual year. In order to provide actuaries and decision makers with a consistent lens through which to view results, rather than using random values generated on-the-fly, a pre-simulated Year-Event-Loss Table (YELT) containing between several thousand and millions of alternative views of a single contractual year is used. The output of aggregate analysis is a Year-Loss Table (YLT). The YELT is generally 1000 times smaller than the YELLT and 1000 times bigger than the YLT. Owing to the large size of data that needs to be processed a weekly simulation can be performed with limited possibility for a real-time simulation.

Traditional database management techniques do not fit the requirements of this stage as data needs to be scanned over rather than randomly access data. Two alternate approaches include accumulation of large memory and accumulation of large distributed file space using high-performance computing seem to be the way forward. While clusters may be efficient for production systems managing risks, ad hoc development and investigations can rely on cloud computing. Methods for accumulating large shared memory includes the use of many-core GPUs for simulating portfolio analysis \cite{agganalysis:conference} which are 15x times faster than the sequential counterpart. The management of large data in memory employs the notion of chunking, which is utilising shared and constant memory as much as possible. A 1 million trial aggregate simulation on a typical contract only takes 25 seconds and can therefore support real-time pricing. Another direction to progress whereby large distributed file space is accumulated will include relying on MapReduce or Hadoop style computations on the cloud.

The last step of the risk analytics pipeline of a modern reinsurance company is referred to as Dynamic Financial Analysis (DFA) \cite{cat8:book}. The aggregate YLTs of catastrophe risks are integrated with investment, reserving, interest rate, market cycle, counter-party, and operational risks in the simulation. The challenge here comes from the combination of YLTs representing different risks which easily results in terabytes of data. From a YLT, a reinsurer can derive important portfolio risk metrics such as the Probable Maximum Loss (PML) \cite{pml:journal} and the Tail Value at Risk (TVAR) \cite{tvar:journal} which are used for both internal risk management and reporting to regulators and rating agencies. Furthermore, these metrics then flow into the final stage in the risk analysis pipeline, namely Enterprise Risk Management, where liability, asset, and other forms of risks are combined and correlated to generate an enterprise wide view of risk.

The challenge lies in the underlying computational complexity and the data size for integrating risks. Similar to the second stage, traditional database management techniques do not fit the requirements of this stage as data needs to be scanned over. Owing to the large size of data pre-computation techniques such as in parallel data warehousing can be applied. Methods in which large distributed file space and memory are accumulated will prove useful.

One characteristic of the reinsurance risk analytics problem is the sudden burst of data in the pipeline. While in the first stage less than ten processors may be sufficient to handle the data, in the second and third stages thousands or even tens of thousands of processors need to be put together to manage and analyse the data. The elastic demand for the storage of data, data retrieval, data processing and data integration makes cloud-based computing attractive.

\section{Conclusions}
\label{conclusions}

In many modern businesses risk analytics plays an important roll, but in reinsurance it is the very core of the enterprise. There are no widgets to be sold, only premium to be exchanged for financial risk based modelled losses. At the very heart of reinsurance is an analytical pipeline based on large scale Monte Carlo simulations. The more data you can analyse and the more simulation trials you can run the better you can manage your aggregate risk, reducing earnings volatility and increasing profit. High performance computing in the form of large distributed memory machines offer great promise in tackling both the computational and data scale challenges posed by reinsurance analtyics. However traditional databases, in our experience, are not a central part of the solution. Rather HPC resources are better suited to either (i) accumulate large quantities of physical memory to support in-memory analytics on large but not enormous datasets less than 1TB) or (ii) to support enormous distributed file systems. Using the first approach one can build highly optimized simulations for core analytical tasks that must be run frequently or even in a `real-time' settings. Using the second approach one can build rich simulation environments that support ad-hoc analytical investigation of truly massive datasets. Perhaps surprisingly, traditional databases don't seem to play a large roll in scaling up reinsurance analytics in an HPC context.

\end{document}